\documentclass[onecolumn,prl, reprint, superscriptaddress]{revtex4-2}
\usepackage{bm}
\usepackage[colorlinks=true,linkcolor=blue,citecolor=blue]{hyperref}
\usepackage{natbib}
\usepackage{times}
\usepackage{amsmath}
\usepackage{amssymb}
\usepackage{amsthm}
\usepackage{amsfonts}
\usepackage{enumerate}
\usepackage{latexsym}
\usepackage{ifpdf}
\newcommand{\beq}{\begin{equation}}
\newcommand{\eeq}{\end{equation}}
\usepackage{graphicx}
\usepackage{makeidx}
\usepackage{color}

\usepackage[version=4]{mhchem}
\usepackage{textcomp,mathcomp}

\begin{document}

\title{Charge transfer and competing symmetry breaking drive orbital reconstruction and emergent ferromagnetism in insulating oxide superlattices}

\author {Nandana Bhattacharya}
\email{nandanab@iisc.ac.in}
\affiliation  {Department of Physics, Indian Institute of Science, Bengaluru  560012, India}

\author {Ranjan Kumar Patel}
\affiliation{Department of Physics, Indian Institute of Science, Bengaluru  560012, India}

\author {Siddharth Kumar}
\affiliation  {Department of Physics, Indian Institute of Science, Bengaluru  560012, India}

\author {Sourav Chowdhury}
\affiliation{Deutsches Elektronen-Synchrotron DESY, 22607 Hamburg, Germany}

\author {Manav Beniwal}
\affiliation{Department of Physics, Indian Institute of Science, Bengaluru 560012, India}

\author {Suresh Chandra Joshi}
\affiliation  {Department of Physics, Indian Institute of Science, Bengaluru  560012, India}

\author {Prithwijit Mandal}
\affiliation  {Department of Physics, Indian Institute of Science, Bengaluru 560012, India}

\author {Jayjit Kumar Dey}
\affiliation{Deutsches Elektronen-Synchrotron DESY, 22607 Hamburg, Germany}

\author {Weibin Li}
\affiliation{ALBA Synchrotron Light Source,
Cerdanyola del Vall\`es, Barcelona E-08290, Spain}

\author {Manuel Valvidares}
\affiliation{ALBA Synchrotron Light Source,
Cerdanyola del Vall\`es, Barcelona E-08290, Spain}

\author{Zhan Zhang}
\affiliation {Advanced Photon Source, Argonne National Laboratory, Lemont, Illinois 60439, USA}

\author{Hua Zhou}
\affiliation {Advanced Photon Source, Argonne National Laboratory, Lemont, Illinois 60439, USA}

\author {Andrei Gloskovskii}
\affiliation{Deutsches Elektronen-Synchrotron DESY, 22607 Hamburg, Germany}

\author {Christoph Schlueter}
\affiliation{Deutsches Elektronen-Synchrotron DESY, 22607 Hamburg, Germany}

\author {Christoph Klewe}
\affiliation{Advanced Light Source, Lawrence Berkeley National Laboratory, Berkeley, CA 94720, USA}

\author {Srimanta Middey}
\email{smiddey@iisc.ac.in}
\affiliation  {Department of Physics, Indian Institute of Science, Bengaluru 560012, India}

\begin{abstract} 

Electron correlation, hopping, and ligand-to-metal charge transfer collectively lead to diverse electronic and magnetic phenomena in 3$d$ transition-metal oxides, where directional $d$ orbitals make hopping highly sensitive to symmetry-dependent orbital overlap. Heterostructure engineering with atomically flat interfaces adds symmetry-breaking charge transfer as a further route to emergent behavior, yet whether interfacial mismatch between constituent oxides of a superlattice shapes ground states independent of epitaxial strain remains unresolved.  Here we examine superlattices combining NdNiO$_3$ with Mott-insulating NdMnO$_3$. Varying layer thickness and combining transport with X-ray spectroscopy, we show that electron transfer from NdMnO$_3$ to NdNiO$_3$ drives a room-temperature insulating state with a distinct electronic structure, accompanied by a reversal in orbital symmetry beyond simple strain considerations, underscoring the interface's central role. These reconstructions stabilize an emergent ferromagnetic insulating state arising from interfacial Ni$^{2+}$-O-Mn$^{4+}$ superexchange. Our results establish a pathway to interface-engineered ferromagnetic insulating phases via competing interactions, with potential for spin-insulatronic applications.

\end{abstract}

\maketitle

\section{Introduction}
Tailoring charge redistribution across interfaces between dissimilar electronic materials has emerged as a central theme in modern condensed matter physics~\cite{Ando:1982p437,Novoselov:2016p9439,Chakhalian:2014p1189}. While charge transfer at conventional semiconductor junctions is largely governed by equilibration of the electron chemical potential~\cite{Kroemer:2001p783}, quantum oxide heterostructures present a far richer landscape owing to the strong coupling among spin, charge, lattice, and orbital degrees of freedom~\cite{Zubko:2011p141}. A canonical example is the polar-catastrophe–driven charge transfer at the interface between the band insulators LaAlO$_3$ and SrTiO$_3$, which gives rise to an emergent two-dimensional electron gas hosting phenomena such as superconductivity, ferromagnetism, Rashba spin–orbit coupling, etc.~\cite{Ohtomo:2004p423,Pai:2018p036503, Ojha:2026arXiv}.
The scenario becomes even more complex when the constituent oxides themselves host strong electron correlations, where the electronic structure cannot be described within a rigid-band picture or treated within independent-electron approximations~\cite{Tsymbal:2012book,Zubko:2011p141}. In several systems, interfacial charge transfer can arise not only from polar discontinuities but also from mechanisms such as nominal valence mismatch and differences in electronegativities [See Ref.~\cite{Chen:2017p243001} for a review]. These processes can stabilize a variety of emergent interfacial phases, including interfacial magnetism~\cite{Bhattacharya:2008p257203,Takahashi:2001p1324}, exchange bias~\cite{Gibert:2012p195}, multiferroicity~\cite{Guo:2017p5062}, superconductivity~\cite{Gozar:2008p782}, etc.
In the case of valence mismatch with a common transition-metal species, such as LaMn$^{+3}$O$_3$/SrMn$^{+4}$O$_3$, LaTi$^{+3}$O$_3$/SrTi$^{+4}$O$_3$~\cite{Yamada:2006p052506,Biscaras:2013p542}, charge redistribution is relatively straightforward and typically occurs over Thomas-Fermi screening length scales ($\sim$ 1–3 unit cells (uc))~\cite{Bhattacharya:2014p65}. In contrast, electronegativity-driven charge transfer across $B$–O–$B'$ bonds presents a far more complex and less predictable scenario~\cite{Bhattacharya:2014p65}. For instance, at SrVO$_3$/SrTiO$_3$ interfaces, although simple band-alignment considerations predict electron transfer, experiments observe no charge transfer into SrTiO$_3$~\cite{Yoshimatsu:2010p147601,Chen:2017p243001}. In principle, owing to the continuity of the oxygen sublattice across the interface, the relative position of the O $2p$ band center ($\epsilon_p$) with respect to the Fermi level ($E_F$) provides a useful descriptor for anticipating charge redistribution~\cite{Chen:2017p243001,Zhong:2017p011023}. Within this framework, electrons are expected to flow from compounds with more negative ($\epsilon_p - E_F$) values to those with less negative ones, consistent with observations in several other interfaces such as LaTiO$_3$/LaNiO$_3$, LaTiO$_3$/LaFeO$_3$, LaMnO$_3$/LaNiO$_3$, LaNiO$_3$/SrIrO$_3$, etc~\cite{Cao:2016p10418,Kleibeuker:2014p237402,Gibert:2015p7355, Liu:2019p19863}. Nevertheless, these systems remain highly entangled: although the direction of charge transfer may often be anticipated, its magnitude and the resulting consequences for the coupled spin, lattice, and orbital degrees of freedom are far more difficult to predict.

Moving beyond single interfaces, the consequences of charge transfer becomes even richer in superlattice architectures, which introduce an additional length scale set by the superlattice periodicity~\cite{Chakhalian:2014p1189,May:2009p892,Ramesh:2019p257}. Depending on the relative magnitudes of the charge-transfer length scale and the superlattice periodicity, the resulting electronic state can evolve from spatially uniform to strongly modulated across the layers~\cite{Tsymbal:2012book}. Most importantly, such digital superlattices can show improved collective properties beyond the randomly doped counterparts~\cite{May:2009p892,Perucchi:2010p4819, Wrobel:2018p035001}. While the periodic interfaces formed between successive $AB$O$_3$/$AB'$O$_3$ layers constitute the primary center of charge redistribution in these superlattices, an additional interface is inevitably present between the film and the underlying substrate [Fig.~\ref{Fig1}(a)]. The latter introduces epitaxial strain, which can modify orbital symmetry and the associated electron hopping and exchange pathways, thereby influencing the emergent phases and creating a classic `chicken-and-egg' interplay among strain, charge transfer, orbital reconstruction and the resulting ground states~\cite{Hwang:2012p103,Middey:2016p305,Martin:2016p16087}. Along this, a critical question remained unanswered: how do these two distinct interfacial influences - the internal $AB$O$_3$/$AB'$O$_3$ interfaces and the film–substrate interface,  collectively shape the emergent physical properties of such systems, especially against the backdrop of strong electronic correlations?

Toward this goal, we focus on $n$ uc NdNiO$_3$/$n$ uc NdMnO$_3$ superlattices, grown under tensile strain [uc = unit cell in pseudocubic setting].
Bulk NdNiO$_3$ (NNO) is a negative charge-transfer system exhibiting a metal–insulator transition near $\sim$ 200 K with $E'$-type antiferromagnetism~\cite{Middey:2016p305}, while NdMnO$_3$ (NMO) is an $A$-type antiferromagnetic Mott insulator with $T_\mathrm{N}\sim80$ K~\cite{Kimura:2003p060403}. Combining electrical transport with element-specific spectroscopic measurements, we use  $n$ uc NNO / $n$ uc NMO  [$n$= 2, 4, 6] platform to directly probe interfacial charge transfer in the presence of both internal Ni/Mn interfaces and the film–substrate interface. Here are our key results:

\begin{enumerate}

\item Interfacial electron transfer from NMO to NNO, leading to an insulating state at room temperature for $n=2$, and 4 superlattices. 
The charge-transfer-driven band reconstruction is fundamentally distinct from  NMO and also oxygen-vacancy-induced charge doping  with comparable degree of electron doping.

\item While the orbital symmetry for the $n=6$ superlattice follows the expectation of an underlying tensile strain, the $n=2$ superlattice exhibit an orbital reversal, dictated by the interfacial $\sigma$-type bonding overall between Ni and Mn.

\item Observation of a rare insulating ferromagnetic state arising due to interfacial electron transfer and orbital reconstruction.

\end{enumerate}

\begin{figure*}[t!] 
	\vspace{-2pt}
	\hspace{-2pt}
	\includegraphics[width=0.99\textwidth]{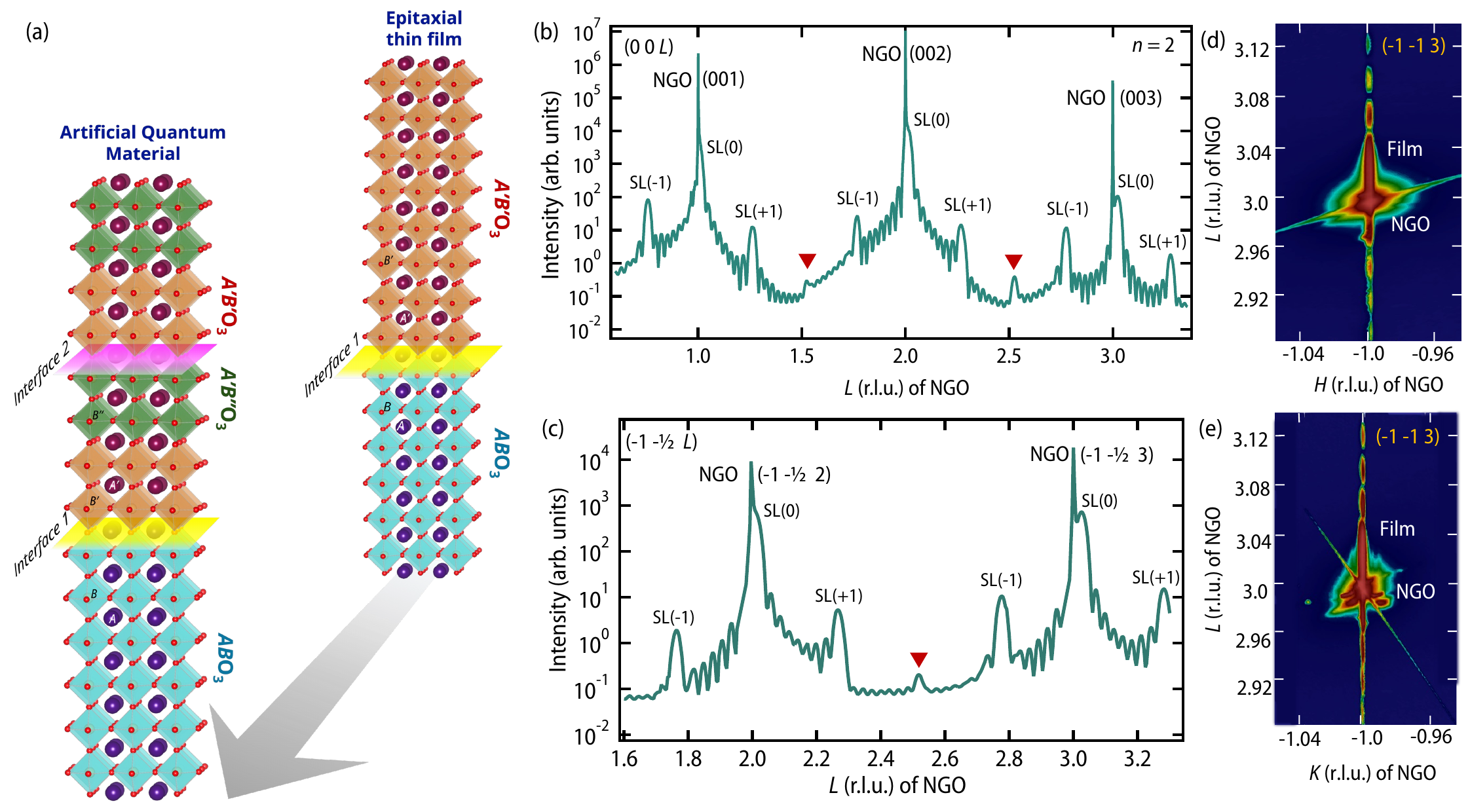}
	\caption{\label{Fig1} {\label{Fig1}\textbf{Thin film characterization:} (a) Schematic comparison between a single-interface $A'B'$O$_3$/$AB$O$_3$ heterostructure and an $A'B'$O$_3$/$AB''$O$_3$ superlattice grown on $AB$O$_3$, illustrating the introduction of an additional interface per periodicity in the latter. (b) Synchrotron X-ray diffraction scan for the $n = 2$ superlattice along the (0 0 $L$) Bragg direction. The  film peak is labeled as SL(0), while the first-order superlattice satellite is denoted as SL($\pm$ 1). Second-order satellite reflections are also observed and marked by red inverted triangles. (c) Off-specular half-order XRD measured along the ($-1$, $-1/2$, $L$) Bragg rod, confirming the same superlattice periodicity as observed in the specular direction. Reciprocal space mapping for the $n = 2$ superlattice around the ($-1$, $-1$, $3$) direction in the (d) $H$–$L$ plane and (e) $K$–$L$ plane.
    }} 
\end{figure*}

\section{Results}

\subsection{Sample growth and characterizations} [$n$ uc NNO / $n$ uc NMO] $\times$ $m$ superlattices ($n = 2, 4, 6$; $m = 9, 4, 3$, respectively) were grown on single terminated NdGaO$_3$ (001)$_\mathrm{pc}$  substrates [$\mathrm{pc}$ denotes pseudocubic setting] using a pulsed laser deposition (PLD) system, connected with a high pressure reflection high-energy electron diffraction (RHEED) setup [growth parameter details are in the experimental section].  RHEED oscillations tracked deposition of each NNO and NMO unit cell, confirming layer-by-layer growth, while streaky RHEED patterns throughout the process verified the retention of smooth surface morphology at all interfaces. These superlattices were subsequently characterized by atomic force microscopy (AFM) and by lab- and synchrotron-based X-ray diffraction (XRD) [details are provided in the experimental section].  The observation of terraces in the AFM image for the $n$ = 2 superlattice further establishes the excellent morphological quality. 

Since this study centers on interface-driven phenomena, and every layer is effectively at an interface for the $n=2$ superlattice, we carried out a detailed synchrotron XRD characterization of this film. The specular (00$L$) scan [Fig.~\ref{Fig1}(b)] shows a broad film peak with Laue fringes, confirming high crystalline quality, and pronounced symmetric first- and second-order superlattice satellites, consistent with the intended 4 uc periodicity. 
A key parameter for interfacial connectivity is the oxygen octahedral rotation (OOR) pattern, described by Glazer notation (
\(+\), \(-\), and \(0\), for in-phase, out-of-phase, and no rotation, respectively)~\cite{Glazer:1972p3384}. Bulk NNO, NMO and NGO adopt an orthorhombic \(a^-b^+c^-\) OOR pattern. Our measurements of a series of half-order Bragg diffraction measurements find \(a^-b^+c^-\) pattern for the $n=2$ superlattice. Although the OOR pattern remains the same across the superlattice, the degree of rotation, and accompanying $A$-site antiparallel displacement may still be distinct for the NNO and NMO layers. To further probe this, we measure the off-specular Bragg rod \((-1, -\tfrac{1}{2}, L)\) [Fig.~\ref{Fig1}(c)]. The scan shows clear superlattice satellite reflections with the same periodicity as the (0 0 \(L\)) scan, confirming that the modulation in the octahedral rotation is spatially periodic across the superlattice. This demonstrates the film’s high structural quality, with sharp interfaces that preserve distinct OOR environments. Since all the superlattices  were grown under the same temperature and pressure conditions, a similar degree of interfacial sharpness is expected across all samples.

Our reciprocal space mapping around the $(-1, -1, 3)$ reflection [Fig.~\ref{Fig1}(d, e)] reveals that the film peak is aligned with the NGO substrate peak in both $H$ and $K$ directions, demonstrating that the superlattice remains coherently strained onto the substrate. All together, these structural characterization results confirm the realization of a structurally coherent, epitaxially locked artificial superlattice with well-controlled periodicity and high interface sharpness.

\subsection{Modification of electronic properties due to superlattice architecture}
Temperature-dependent electrical transport measurements, carried out in Van der Pauw geometry, (see Experimental section for details) reveal a systematic evolution with layer periodicity $n$ [Fig.~\ref{Fig2}(a)].  The pristine NNO film exhibits a sharp MIT near $\sim 160$ K, accompanied by pronounced thermal hysteresis, characteristic of a first-order transition~\cite{Bhattacharya:2025p2418490}. This origin of this MIT is attributed to a bond disproportionation transition~\cite{Park:2012p156402,Bisogni:2016p13017,Middey:2018p156801}. 
For the $n=6$ superlattice, the $T_\mathrm{MIT}$ is suppressed to $\sim 100$  K, with negligible hysteresis, although the overall resistance in the metallic phase remains higher than that of the pristine NNO.
The room-temperature sheet resistance ($R_s$) increases monotonically with decreasing $n$, indicating progressive carrier localization. Both the $n = 4$ and $n = 2$ superlattices  are insulating over the entire measured temperature range; however, the low-temperature enhancement in $R_s$ is substantially weaker for the $n = 4$ film.
The insulating behavior of the $n = 2$ superlattice can be described by an activated transport model, $R_s = R_0 \exp\left(\frac{E_g}{2 k_B T}\right)$, yielding an energy gap of $E_g \sim 147$ meV. Notably, the parallel-resistance model constructed from pristine NNO and the $n=2$ superlattice is unable to capture the resistivity behavior of the $n=4$ superlattice, which can be attributed to the contribution of hole doping in the manganite layer, discussed later.
Furthermore, in contrast to LaNiO$_3$/LaMnO$_3$ superlattices, where weak metallicity persists down to $n \geq 3$ and a fully insulating state emerges only below $n = 2$~\cite{Hoffman:2013p14411}, the insulating phase in the present system is stabilized already at $n = 4$. This earlier onset is indicative of a stronger renormalization of the electronic bandwidth, likely driven by stronger $RE$-dependent distortions of the Ni-O-Ni and Ni-O-Mn bond networks that reduce the effective hopping amplitude.

\begin{figure*}[t!] 
    \centering
	\vspace{-2pt}
	\hspace{-2pt}
	\includegraphics[width=0.85\textwidth]{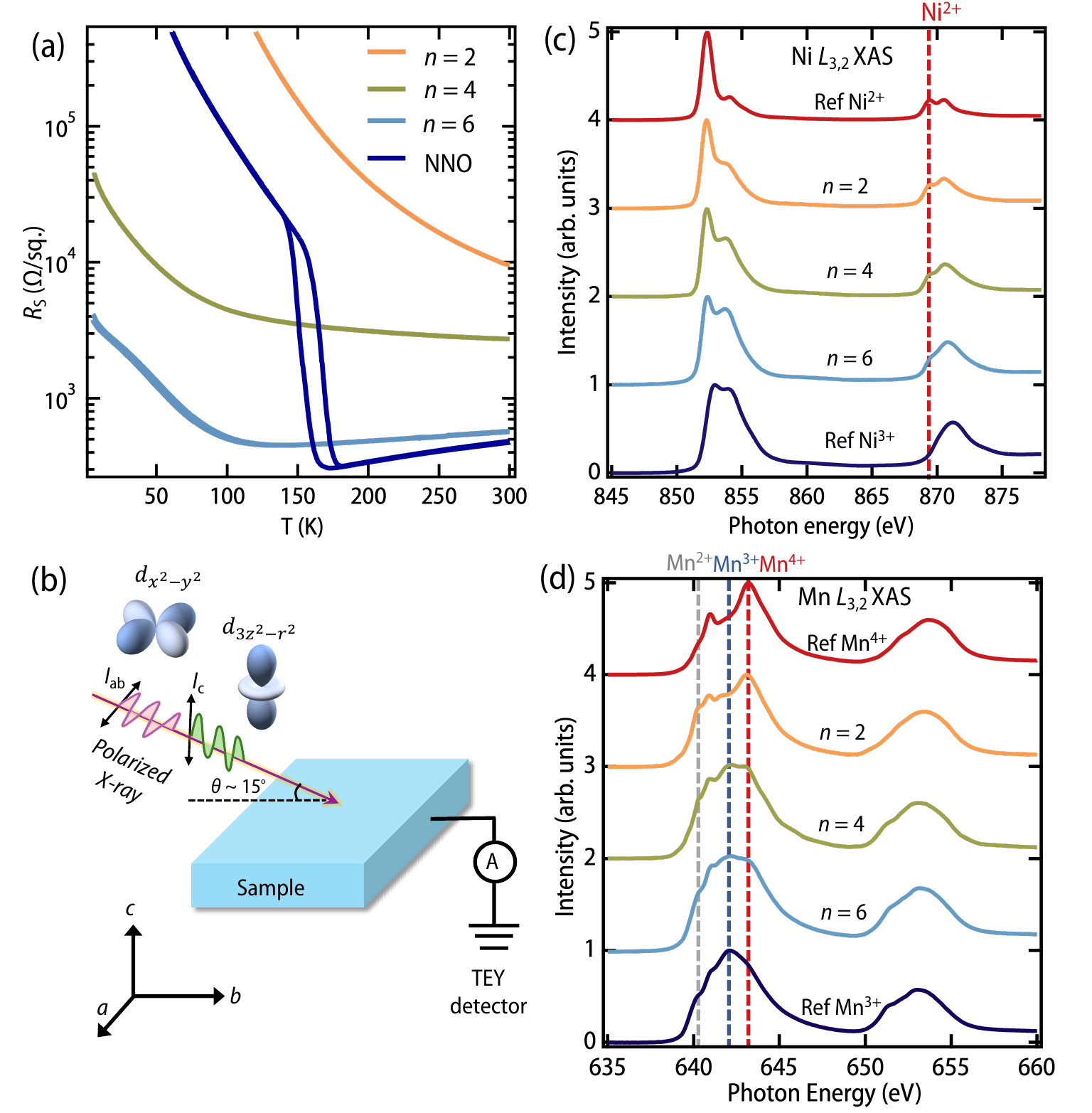}
	\caption{\label{Fig2}{\label{Fig2}\textbf{Electrical transport and demonstration of charge transfer:} (a) Sheet resistance versus temperature plots for  the $n= $ 2, 4 and 6 superlattices along with pristine NNO film [also used in~\cite{Bhattacharya:2025p2418490}]. (b) Schematic depiction of the XAS measurement setup. Linearly polarized X-rays (corresponding absorption spectra $I_\mathrm{c}$ and $I_\mathrm{ab}$ respectively) are incident on the films at grazing incidence, and the spectra are recorded in total electron yield (TEY) mode. The average of these spectra is the recorded XAS at room temperature. (c) Ni $L_{3,2}$ edge XAS for superlattices with $n = 2, 4,$ and $6$. Reference spectra for Ni$^{3+}$ is from 15 uc NNO on NGO [also used in ~\cite{Bhattacharya:2026p130}], and for Ni$^{2+}$ from 20 uc Nd$_2$NiMnO$_6$ grown on NGO adapted from ~\cite{Bhattacharya:2025p176201}. (d) Mn $L_{3,2}$ edge XAS for superlattices  with $n = 2, 4,$ and $6$. Reference spectra for Mn$^{3+}$ is taken from 20 uc NMO on NGO [also used in ~\cite{Bhattacharya:2026p130}], while Mn$^{4+}$ reference is from 20 uc Nd$_2$NiMnO$_6$ on NGO, adapted from ~\cite{Bhattacharya:2025p176201}. The spectra in (c) and (d) have been vertically shifted for clarity.
    }}
\end{figure*}

\subsection{Demonstration of electron transfer between NNO and NMO} To elucidate the electronic structure underlying the observed evolution in transport behavior, we investigate the valence states of Ni and Mn in the superlattices  using X-ray absorption spectroscopy (XAS) at room-temperature. The measurements were performed at the Advanced Light Source (USA) in total electron yield (TEY) mode [see Fig.~\ref{Fig2}(b) for schematic and the experimental section for measurement details]. Spectra were collected at the Ni and Mn $L_{3,2}$ edges, which are highly sensitive to their oxidation states~\cite{De:1994p529}, thereby enabling a direct assessment of interfacial charge transfer. In their bulk forms, pristine NNO and NMO host Ni and Mn with $3+$ valence state. Fig.~\ref{Fig2}(c) presents the Ni $L_{3,2}$-edge spectra for the superlattices, along with reference spectra corresponding to Ni$^{3+}$ and Ni$^{2+}$ in an octahedral coordination environment. The superlattice spectra reveal a clear coexistence of Ni$^{2+}$ and Ni$^{3+}$ states, with a systematic increase in Ni$^{2+}$ character as the layer number $n$ is reduced, implying substantial electron transfer to the Ni sites. This trend reflects the increasing dominance of interfacial effects at reduced thickness, where the electronic structure is governed largely by the interface-driven charge redistribution. Concomitantly, the Mn $L_{3,2}$-edge spectra display mixed Mn$^{3+}$/Mn$^{4+}$ character, with the Mn$^{4+}$ contribution most pronounced for $n = 2$ film, consistent with enhanced charge transfer from Mn to Ni in the interface-dominated limit [Fig.~\ref{Fig2}(d)]. Additionally, a weak feature near $\sim 640$ eV is observed in all films, which is attributed to surface Mn$^{2+}$ states, very common in  manganite thin films due to surface symmetry breaking~\cite{Pesquera:2012p1189,Bhattacharya:2025p235438,Tebano:2008p137401}. 

To estimate the amount of electron transfer, the Ni $L_2$-edge spectra were fitted using a linear combination of Ni$^{2+}$ and Ni$^{3+}$ reference spectra. The extracted Ni$^{2+}$ percentages for $n = 6, 4,$ and $2$ were found to be 12\%, 24\%, and 48\%, respectively, exhibiting a systematic increase with decreasing $n$. 
This suggests that, for the $n=2$ superlattice, there is an average transfer of approximately 0.5 electron to each Ni site, as every layer lies at an interface. Taking into account the TEY probing depth ($\sim$ 6 nm), we have also estimated the number of active interfaces that predominantly contribute to the XAS signal for the $n$ = 4, 6 superlattice. Interestingly, these considerations consistently reproduce the observed Ni$^{2+}$ fraction in the $n$ = 4, 6 superlattices. Overall, this implies that the charge transfer is predominantly interfacial in nature and confined within $\sim 1$–2 uc across the interface.
 Similar XAS analysis is difficult for Mn edge due to much complex multiplet features in mix-valent system. Hence, we independently assess the Mn valence using Mn 3$s$ core level hard X-ray photoelectron spectroscopy (HAXPES) at room temperature, performed at DESY, Germany [see experimental section]. The exchange splitting $\Delta E_{3s}$ provides an estimate of the average Mn oxidation state~\cite{Galakhov:2002p113102}. For the $n = 2$ superlattice, we obtain $\Delta E_{3s} \sim 4.6$ eV, corresponding to an average Mn valence of $\sim + 3.6$. 
Relative to the nominal Mn$^{3+}$ state in NMO, this indicates a charge transfer of about 0.6 electrons per Mn site. This value is in close agreement with that expected from electron donation to the NNO layers in $n=2$ superlattice, which exhibit an average Ni oxidation state of +2.5.
 
\begin{figure*}[t!] 
	\vspace{-2pt}
	\hspace{-2pt}
	\includegraphics[width=1.0\textwidth]{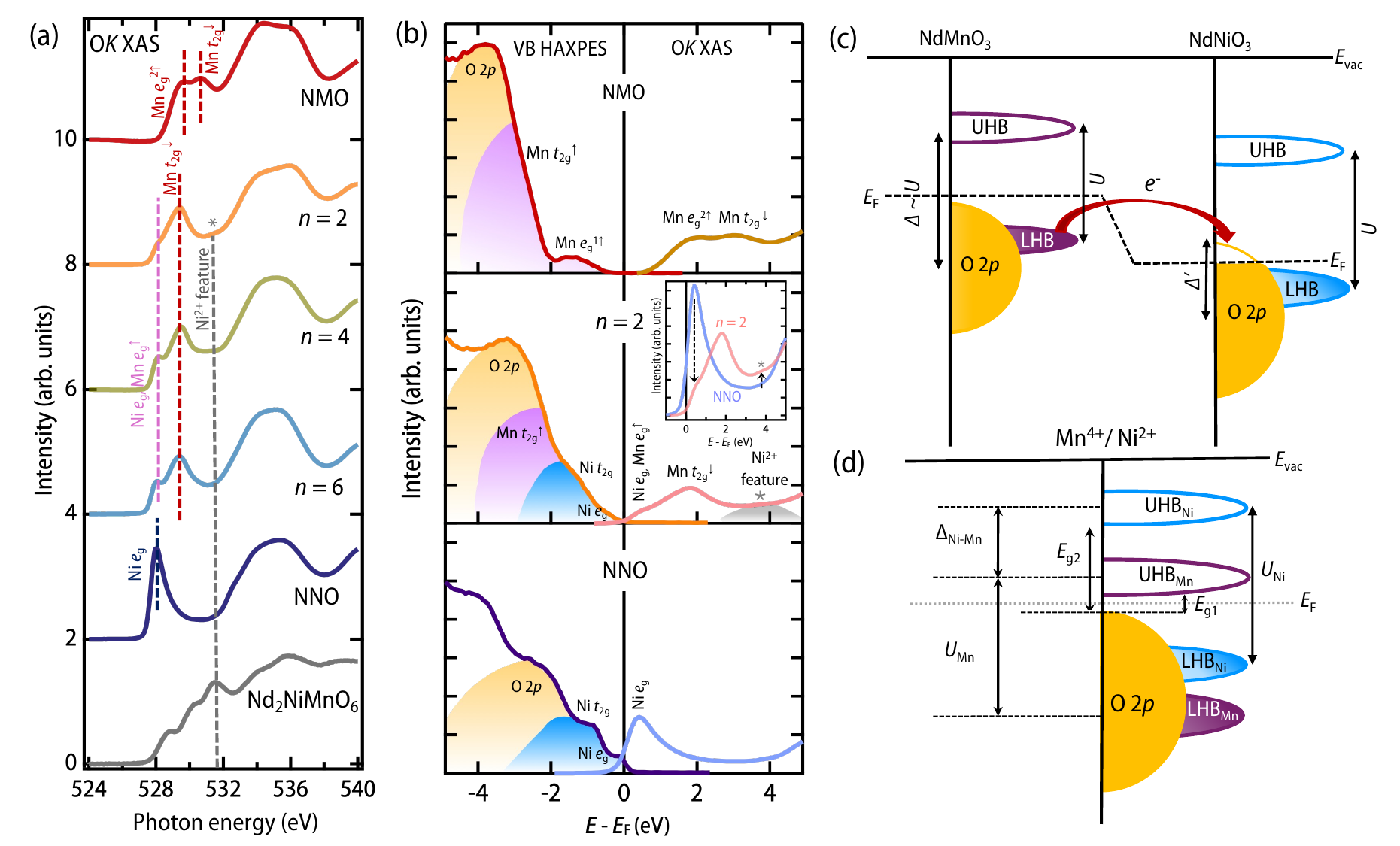}
	\caption{\label{Fig3} {\label{Fig3}\textbf{Correlated band reconstruction on interfacial charge transfer:} (a) O $K$-edge XAS for pristine 15 uc NNO, 20 uc NMO, 20 uc Nd$_2$NiMnO$_6$ [for reference of Ni$^{2+}$ states reflected in O $K$ edge~\cite{Bhattacharya:2025p176201}] films and superlattices  $n$ = 2,4 and 6 films on NGO. The spectra have been vertically shifted for clarity. (b) Electronic states near the Fermi energy ($E_F$) using VB HAXPES (for occupied states) and O $K$-edge XAS (for unoccupied states) for: NMO (top), $n=2$ superlattice (centre) and NNO (bottom). Inset shows the suppression of prepeak in $n=2$ superlattice compared to pristine NNO, and enhancement of spectral weight $\sim$ 541.5 eV highlighting emergence of Ni$^{2+}$ states. Shaded regions denote the occupied LHB states (c) Schematic representation under the Mott-Hubbard scheme for NMO, negative-charge-transfer scheme for NNO, depicting the charge transfer. (d) Correlated band reconstruction for $n=2$ superlattice derived from combined VB and O $K$-edge XAS. Two distinct band gaps $E_{g1}$ and $E_{g2}$ along with charge transfer energy scale $\Delta_\mathrm{Ni-Mn}$ have been marked. The UHB, LHB and $U$ parameter for individual Ni and Mn have been marked in subscript.
    }}
\end{figure*} 

\subsection{Interfacial correlated band reconstruction} We note that in oxygen-deficient $RE$NiO$_{2.5}$, with an Ni valence change comparable to what we found for $n=2$ superlattice, gives rise to a $\sim2$ eV gap that defines its insulating behavior~\cite{Kotiuga:2019p115002}. The much smaller transport gap inferred above for the $n=2$ superlattice therefore motivates an investigation of how its correlated electronic structure evolves under a similar degree of charge transfer. To examine this, we combine O $K$-edge XAS, which probes the unoccupied density of states above $E_F$, with valence-band (VB) HAXPES to access the occupied states at room temperature.

For reference NNO film, a pre-peak in the O $K$ edge at $\sim 527$–$528$ eV [Fig.~\ref{Fig3}(a)] arises from the $3d^{8}\underline{L} \rightarrow \underline{c}3d^{8}$ transition (where $\underline{c}$ denotes the O $1s$ core hole). The intensity of this feature serves as a direct measure of Ni–O covalency~\cite{Middey:2016p305}. For the reference NMO, spectral features in the $\sim 528.5$ – $531$ eV range arise from Mn $3d$–O $2p$ hybridized states. The Mn$^{3+}$ ions here are Jahn–Teller active, leading to a distortion of the oxygen octahedra that lifts the degeneracy of the $e_g$ orbitals, resulting in an occupied $e_g^{1\uparrow}$ state below $E_F$ and an unoccupied $e_g^{2\uparrow}$ state above $E_F$ [Fig.~\ref{Fig3}(a)] ~\cite{Yu:2010p027201}.

As shown in Fig.~\ref{Fig3}(a), the pre-peak intensity is progressively suppressed with decreasing $n$ in superlattice geometry, relative to pristine NNO, as expected for electron transfer into the Ni sites and the filling of ligand holes on oxygen [See inset ~\ref{Fig3}(b) for comparison between $n=2$ superlattice and NNO]. 
Furthermore, a transfer of spectral weight towards higher photon energies ($\sim 531.5$ eV), corresponding to the emergence of Ni$^{2+}$ states~\cite{Kuiper:1989p221, Bhattacharya:2025p176201}, is also observed. Such evolution signifies a gradual crossover from a $3d^{8}\underline{L}$-dominated configuration toward increased $3d^{8}$ character~\cite{Cao:2016p10418}. Furthermore, the Mn $3d$–O $2p$-derived states above $E_F$ systematically shift toward lower photon energies with decreasing $n$. Notably, the $e_g^{2\uparrow}$ feature evolves toward a predominantly single $e_g^{\uparrow}$ character, indicating an increase in Mn oxidation state toward Mn$^{4+}$~\cite{Saitoh:1995p13942,Galdi:2012p125129}, and corroborates interfacial charge transfer from Mn to Ni.

We now turn to our observation of VB HAXPES spectra. For NNO [Fig.~\ref{Fig3}(b), lower panel],  the spectral weight near the Fermi level ($E_F$) is dominated by Ni-derived states (blue shaded region), originating from the filled $t_{2g}$ manifold and partially occupied $e_g$ states, both strongly hybridized with O $2p$ orbitals~\cite{Barman:1994p8475}. The finite intensity at $E = E_F$, arising from these hybridized $e_g$ states, is consistent with the metallic character of NNO at room temperature. The broader O $2p$ states (yellow shaded region) appear at higher binding energies, extending beyond $\sim 2$ eV. Subsequently, for NMO, the occupied Mn $d$ states below $E_F$ consist of a lower-energy feature corresponding to the $e_g^{1\uparrow}$ state, arising from Jahn-Teller distortion, and higher-energy filled $t_{2g}^{\uparrow}$ states (pink shaded regions) [Fig.~\ref{Fig3}(b), upper panel]. There is no observable spectral weight at $E=E_F$, consistent with its insulating nature [Fig.~\ref{Fig3}(b), upper panel]. The feature centered around $\sim 3-4$ eV below Fermi, extending over a broad energy range (yellow shaded region), corresponds to the O $2p$ states. 
In case of the $n = 2$ superlattice [Fig.~\ref{Fig3}(b), center panel], the states near $E_F$ are primarily derived from Ni $t_{2g}$ and $e_g$ orbitals, since with the increase in Mn$^{4+}$ valence, the occupied Mn $e_g^{1\uparrow}$ spectral weight becomes negligible~\cite{Saitoh:1995p13942,Galdi:2012p125129}.

To reconstruct the full correlated electronic structure,  the O $K$-edge spectra are aligned such that the energy separation between the O $2p$ states and the prepeak of NNO is $\sim 2.5$~eV, following prior reports~\cite{Horiba:2007p155104,Hepting:2020p381}. Figure~\ref{Fig3}(c) schematically illustrates the electronic structures of NNO and NMO, constructed from the combined VB and O $K$-edge spectra within the Mott-Hubbard and negative charge-transfer frameworks, respectively, highlighting the direction of interfacial electron flow. Within this framework, we reconstruct the band alignment for the $n = 2$ superlattice by combining the Ni and Mn features, showing two characteristic energy gaps.  The first gap, $E_{g1} \sim 147$ meV [Fig.~\ref{Fig3}(d)], corresponds to the transport gap. 
The second gap arises from the energy separation between the unoccupied Ni and Mn upper Hubbard bands (UHB), defining a charge-transfer energy scale $\Delta_{\mathrm{Ni-Mn}} \sim 2$ eV. This leads to an effective higher-energy gap, $E_{g2} = E_{g1} + \Delta_{\mathrm{Ni-Mn}}$, associated with interfacial excitations between Ni and Mn states.
Such a gap reflects an interface-induced band reconstruction driven by correlated charge redistribution and is distinct from a conventional Mott gap (intra-site $d$-$d$ excitations) or a charge-transfer gap (O 2$p$ to TM 3$d$ excitations), serving as a hallmark of interface-engineered correlated states in superlattices~\cite{Cao:2016p10418,Chen:2013p116403}.

\begin{figure*}[t!] 
	\vspace{-2pt}
	\hspace{-2pt}
	\includegraphics[width=0.95\textwidth]{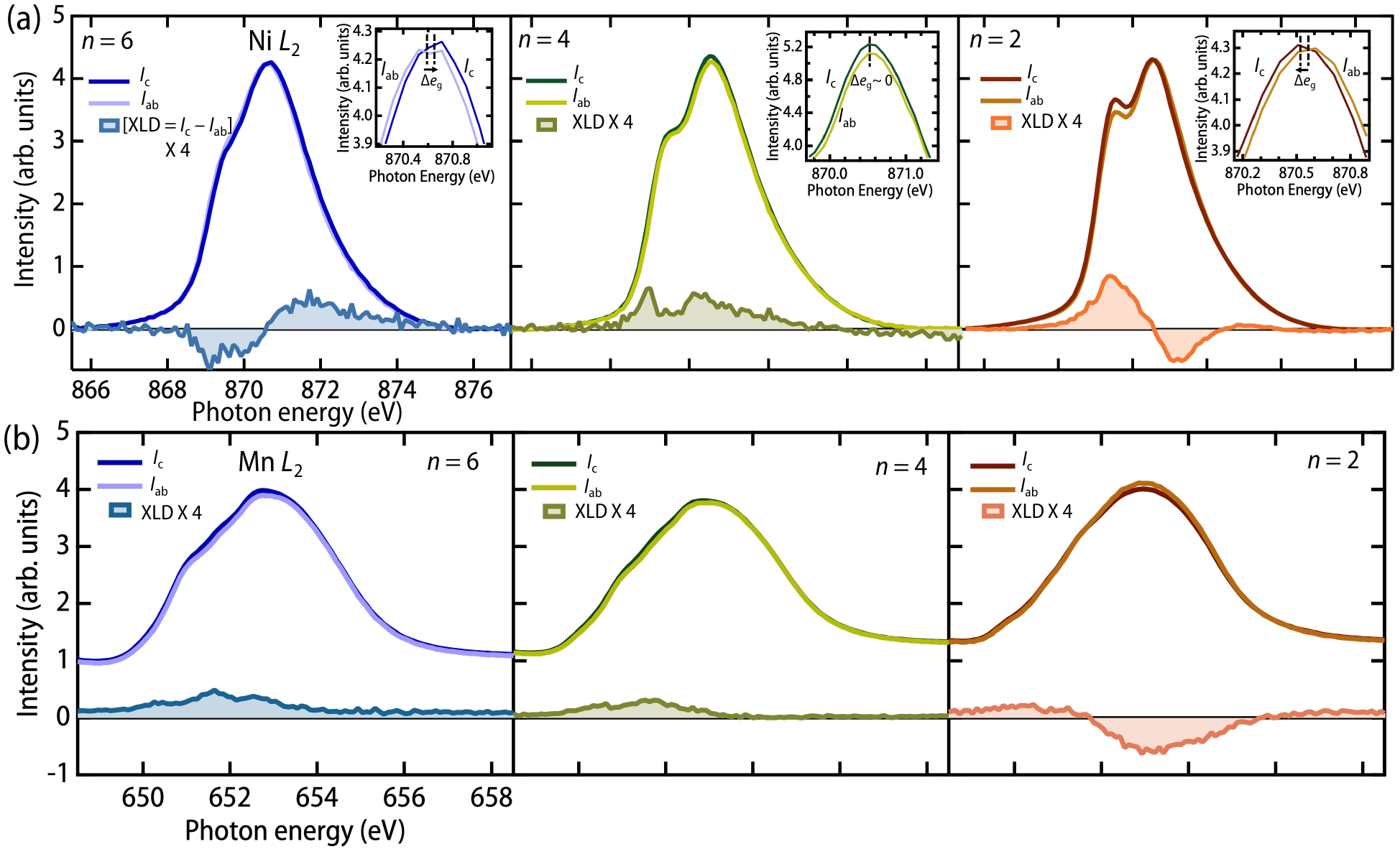}
	\caption{\label{Fig4} {\label{Fig4} \textbf{Orbital reconstruction due to competing interfacial symmetries:}  XAS spectra for $n = 6$, 4, and 2 superlattices  measured with out-of-plane polarized ($I_c$) and in-plane polarized ($I_{ab}$) X-rays. The corresponding XLD ($I_c - I_{ab}$) signals (scaled for clarity) are shown below each spectrum, with insets highlighting the energy splitting $\Delta e_g$ corresponding to the energy shift between the $I_c$ and $I_{ab}$ peak at the (a) Ni $L_2$-edge and (b) Mn $L_2$-edge.
    }}
\end{figure*}

\subsection{Charge transfer induced orbital reconstruction and emergent ferromagnetism}
After confirming that the band reconstruction due to interfacial electron transfer is different compared to heavy electron doping via oxygen vacancy, we focus on the underlying response of the TM orbitals. To probe this, we explore the orbital symmetries in all the superlattices using X-ray linear dichroism (XLD) measurements. The difference in $L_{3,2}$ absorption spectra, measured with out-of-plane polarized X-rays ($I_{c}$) and in-plane polarized X-rays ($I_{ab}$) [See Fig.~\ref{Fig2}(b) for schematic representation], reflects the difference in electron occupancies between out-of-plane ($d_{3z^2-r^2}$, $d_{xz}$, $d_{yz}$)  and in-plane ($d_{x^2-y^2}$, $d_{xy}$) orbitals~\cite{Chakhalian:2007p1114,Mandal:2021p060504,Tebano:2008p137401}. 
Fig.~\ref{Fig4}(a) summarizes the XLD (= $I_{c}$-$I_{ab}$) response at the Ni $L_2$ edge for the superlattices. The XLD spectra for the $n=6$ superlattice exhibits a derivative-like line shape, implying that the $d_{x^2-y^2}$ orbital lies at a lower energy than the $d_{3z^2-r^2}$ orbital. This splitting is consistent with the tensile strain imposed by the substrate, leading to $d_{x^2-y^2}$ states at a lower energy. 
In sharp contrast, the derivative-like XLD signal reverses sign for the $n=2$ superlattice, even though the film is epitaxially strained with the underlying substrate. The lowering of the Ni $d_{3z^2-r^2}$ at the interface is related to the inter-orbital hopping between Ni and Mn, which we have discussed in the later part of the paper. 
At the intermediate thickness ($n = 4$), the derivative-like feature is largely suppressed, and the XLD signal shows a positive value, indicating the electron occupation of $d_{x^2-y^2}$ is higher even though both orbital have similar energy positions. A complementary trend is also observed at the Mn $L_2$ edge [Fig.~\ref{Fig4}(b)]. Here, a positive XLD corresponds to preferential occupation of the $d_{3z^2-r^2}$ orbital, while a negative XLD indicates enhanced occupation of the $d_{x^2-y^2}$ orbital~\cite{Pesquera:2012p1189,Tebano:2008p137401}. Consistent with the behavior observed for Ni, the Mn XLD evolves from predominantly strain-driven in-plane orbital occupation (+ve XLD) in the $n=6$ superlattice to enhanced $d_{3z^2-r^2}$ occupation (-ve XLD) in the ultrathin limit ($n=2$). 

\begin{figure*}[ht!]
	\vspace{-2pt}
	\hspace{-2pt}
	\includegraphics[width=1.0\textwidth]{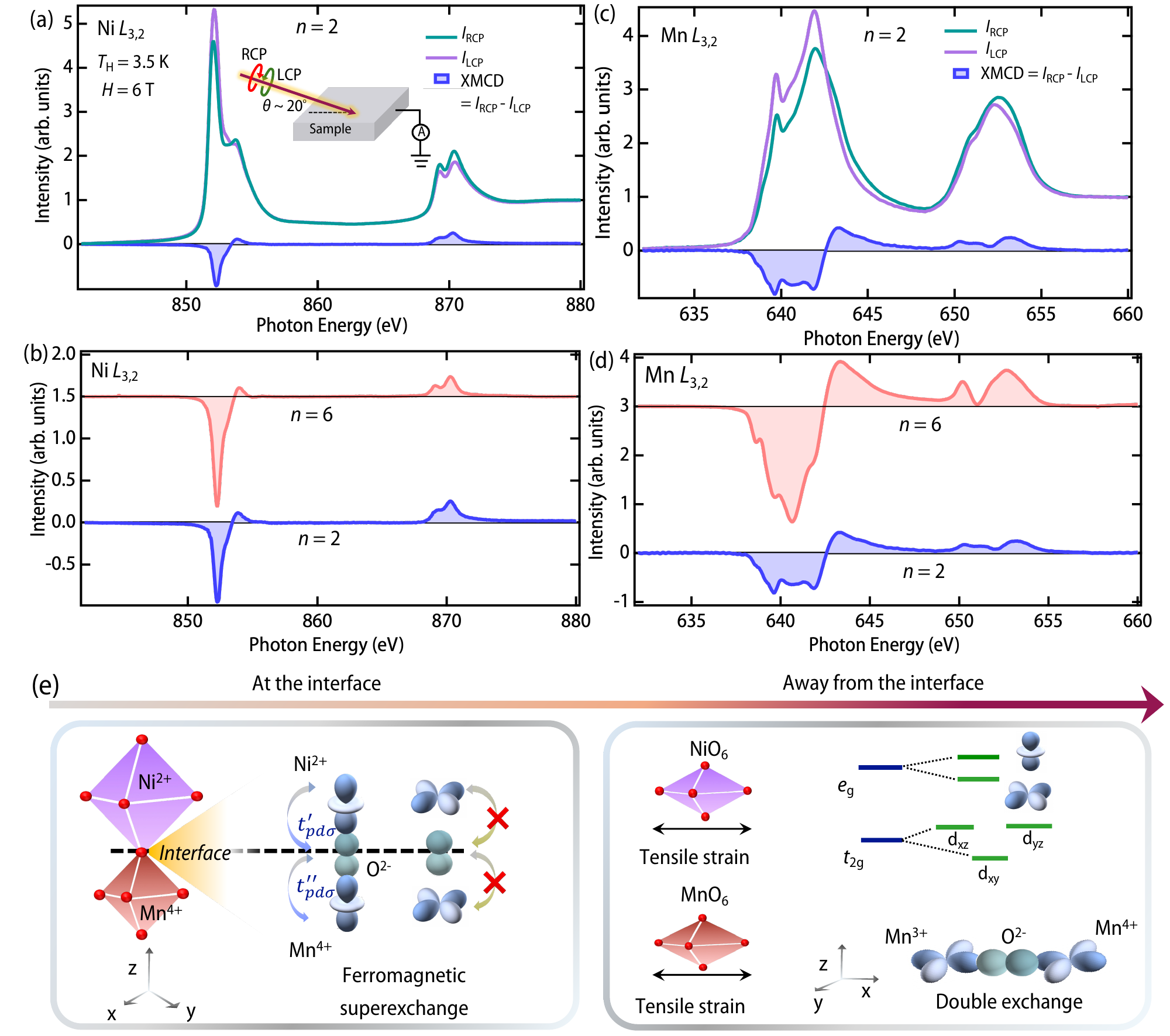}
	\caption{\label{Fig5} {\label{Fig5}\textbf{Emergent ferromagnetism and exchange pathways:} (a) Right ($I_\mathrm{RCP}$) and left ($I_\mathrm{LCP}$) circularly polarized absorption spectra, and XMCD for the $n=2$ superlattice at the Ni $L_{3,2}$ edge. Inset: Schematic of the XMCD measurement geometry. (b) XMCD at Ni $L_{3,2}$ edge for $n=2$ and $n=6$ superlattices. (c) $I_\mathrm{RCP}$, $I_\mathrm{LCP}$, and corresponding XMCD for the $n=2$ superlattice at the Mn $L_{3,2}$ edge. (d) XMCD at Mn $L_{3,2}$ edge for $n=2$ and $n=6$ superlattices. (e) Comprehensive schematic of  orbital symmetry and magnetic exchange. Near the interface, the $d_{3z^2-r^2}$ occupancy is favored due to interfacial hybridization with the hopping parameter $t_{pd\sigma}^{'}$ and $t_{pd\sigma}^{''}$ for Ni and Mn sites respectively. This promotes a ferromagnetic superexchange across the interface.  Away from the interface, the crystal field splitting under tensile strain of the $e_g$ manifolds stabilizes preferential $d_{x^2-y^2}$ orbital occupancy for Ni and Mn states. Here, a double-exchange like behavior is more dominant among the Mn states.
    }}
\end{figure*} 

This thickness-dependent crossover in orbital symmetry highlights a striking competition between the NNO/NMO versus the film/substrate interfaces. In the ultrathin limit ($n=2$), preferential alignment of the $d_{3z^2-r^2}$ orbitals along the out-of-plane Ni-O-Mn bond (direction $[001]$) is observed, which elucidates the dominance of the NNO/NMO interfacial contribution. This bond promotes $\sigma$-type overlap ($t_{pd\sigma}^{'}$ and $t_{pd\sigma}^{''}$ for Ni and Mn sites respectively)~\cite{Slater:1954p1498}, providing an efficient pathway for charge transfer. The $d_{x^2-y^2}$ orbitals have forbidden coupling across this interface, underscoring the strong directional selectivity of the interfacial electronic reconstruction. In contrast, for the superlattice ($n=6$) with a greater fraction of non-interfacial layers, the effect of tensile strain enforced by the substrate dominates over the interfacial orbital reconstruction effect, leading to the observation of $d_{x^2-y^2}$ at a lower energy in XLD measurement.

With the partial interfacial charge transfer and the resulting evolution of orbital occupancies discussed above, several competing magnetic exchange pathways become possible within the superlattices, including Ni$^{2+}$-O-Mn$^{4+}$, Ni$^{3+}$-O-Mn$^{3+}$, Mn$^{3+}$-O-Mn$^{4+}$, Mn$^{3+}$-O-Mn$^{3+}$, and Ni$^{3+}$-O-Ni$^{3+}$ interactions. While the Ni$^{2+}$-O-Mn$^{4+}$ superexchange~\cite{Pal:2019p045122} and Mn$^{3+}$-O-Mn$^{4+}$ double-exchange~\cite{Koide:2001p246404,Goodenough:1955p564} pathways are expected to favor ferromagnetic coupling, the Ni$^{3+}$-O-Mn$^{3+}$, Mn$^{3+}$-O-Mn$^{3+}$, and Ni$^{3+}$-O-Ni$^{3+}$ interactions generally promote antiferromagnetic exchange~\cite{Goodenough:1955p564,Middey:2016p305}.  To investigate the resulting magnetic state, we perform element-specific X-ray magnetic circular dichroism (XMCD) measurements, in which XAS is recorded with left- and right-circularly polarized X-rays [inset of Fig.~\ref{Fig5}], and the difference is directly proportional to the magnetization [see experimental section]. Fig.~\ref{Fig5}(a)-(d) show the Ni and Mn $L_{3,2}$-edge XMCD spectra for the $n=2$ and $n=6$ superlattices  measured at 3.5 K under a magnetic field of 6 T.  The negative sign of the $L_3$-edge of the XMCD signal for both Ni and Mn confirms that the moments associated with these two sublattices are ferromagnetically coupled ~\cite{Van:2014p95} [Fig.~\ref{Fig5}(b),(d)]. Thus, the ferromagnetic exchanges present in the superlattices, are dominant over the antiferromagnetic exchanges mentioned above. While the Ni XMCD line shape remains nearly unchanged between the two superlattices  and closely resembles the characteristic fingerprint of Ni$^{2+}$~\cite{Piamonteze:2015p014426} [Fig.~\ref{Fig5}(b)], the Mn XMCD spectra exhibit pronounced differences, reflecting changes in the relative contributions of Mn$^{3+}$ and Mn$^{4+}$ to the magnetic response [Fig.~\ref{Fig5}(d)]. 

To quantify these differences, XMCD sum-rule analysis was employed. We note that for both the Ni and Mn cases, the orbital moment $m_l$ is quenched, characteristic of 3$d$ transition metal cations, and the dominant contribution is that of $m_s$. For both the $n=2$ and 6 superlattices, we find a  Ni moment $\sim$ 0.4 $\mu_B$/Ni, which is comparable to previous experimental and theoretical reports on LaNiO$_3$/LaMnO$_3$ superlattices  ~\cite{Hoffman:2013p14411,Piamonteze:2015p014426,Lee:2013p035126}.

We next consider the Mn $L_{3,2}$ edge. For the $n=2$ superlattice, the extracted moment is $\sim 1.53 ~\mu_B$/Mn, comparable to experimentally reported values for the ultrathin limit of LaNiO$_3$/LaMnO$_3$ superlattices~\cite{Hoffman:2013p14411}. Importantly, the Mn XMCD line shape closely resembles that commonly observed in $RE_2$Ni$^{2+}$Mn$^{4+}$O$_6$ double perovskites ~\cite{Pal:2019p045122}, indicating that the magnetic ground state is predominantly governed by ferromagnetic superexchange interactions between Ni$^{2+}$ and Mn$^{4+}$~\cite{Das:2008p186402}. As this superlattice is insulating, the  Mn$^{3+}$-O-Mn$^{4+}$ ferromagnetic double-exchange interaction is absent.

For the $n=6$ superlattice, an Mn formal valence of $\sim3.12+$ is inferred from charge conservation. The XMCD analysis yields a Mn moment of $\sim 2.7~\mu_B$, consistent with the range of moments reported for thicker nickelate/manganite superlattices~\cite{Piamonteze:2015p014426}. In contrast to the $n=2$ superlattice, the Mn XMCD line shape  acquires features characteristic of mixed-valence manganites~\cite{Koide:2001p246404}, indicating that the magnetic interactions are no longer governed solely by interfacial Ni$^{2+}$-O-Mn$^{4+}$ superexchange. This behavior suggests a spatial evolution of the Mn valence across the NdMnO$_3$ layer. Near the interface, Mn remains predominantly Mn$^{4+}$ and participates in Ni$^{2+}$-O-Mn$^{4+}$ ferromagnetic superexchange, while regions just away from the interface contain mixed-valence Mn$^{3+}$/Mn$^{4+}$ ions that support Mn$^{3+}$-O-Mn$^{4+}$ double exchange, which also contributes strongly to the observed XMCD line shape. 
Interestingly, we also find that the ferroamagnetic coupling from XMCD is persistent even at higher temperatures of $\sim$ 79 K (see experimental section for measurement details), indicative of a robust ferromagnetic  behavior for all three superlattices. We further note that the presence of different possible antiferromagnetic interactions can account for the reduced Mn moment found, relative to that expected from the theoretical estimates ~\cite{Piamonteze:2015p014426,Lee:2013p035126}.

\section{Conclusions}
In summary, we demonstrate that interfacial charge transfer in NNO/NMO superlattices is governed by a critical length scale and proceeds through a mechanism fundamentally distinct from conventional chemical doping. In the strong charge-transfer limit ($n=2$), a fully reconstructed, correlation-driven insulating state emerges, accompanied by robust ferromagnetic coupling between Ni and Mn mediated by Ni$^{2+}$-O-Mn$^{4+}$ superexchange. Increasing the layer thickness drives a crossover to a mixed-phase regime in $n=4$ and $n=6$, where charge transfer becomes increasingly confined to the interfaces. We further uncover a strain versus interface hybridization-driven reversal of orbital symmetry that promotes interfacial ferromagnetism through favorable superexchange pathways. Away from the interface, however, the emergence of mixed-valence Mn$^{3+}$/Mn$^{4+}$ states activates a double-exchange mechanism that enhances carrier itinerancy and contributes to the reduced low-temperature resistivity observed in the thicker superlattices. Fig.~\ref{Fig5}(e) provides a unified schematic representation of the orbital hybridization and magnetic exchange interactions. 
Due to the presence of both interfacial and bulk-like layers, competing interactions may stabilize non-collinear spin textures, which can be probed by  polarized neutron reflectometry,  x-ray resonant magnetic reflectivity~\cite{Hoffman:2016p041038,Fabbris:2018p180401}.  
Overall, our work provides a clean experimental demonstration of how orbital reconstruction at interfaces can stabilize emergent magnetic phases in insulating systems, with implications for spin-insulatronics and low-dissipation spintronic technologies~\cite{Brataas:2020p1,Li:2025p016702}.

\par
\section{Experimental section}

\textbf{\emph{Thin film growth :}}
Single-terminated NGO substrates were obtained by thermal treatment of as received substrate from Shinkosha, Japan~\cite{Patel:2023p031407}.
$n = 2, 4$ and 6 superlattices  were grown in layer-by-layer fashion on GaO$_2$ single-terminated NGO [110]$_\mathrm{or}$ substrates, using a Neocera PLD system. The KrF excimer laser ($\lambda$ $\sim$ 248 nm) was operated at a repetition rate of 2 Hz and fluence of 2 J/cm$^2$. The films were grown under identical conditions at a substrate temperature of 750 $^{\circ}$C and a dynamic oxygen pressure of 150 mTorr. Post-growth annealing was carried out at the deposition temperature in 500 Torr of oxygen pressure. The unit cell precision within each stack of the superlattice was monitored using an in-situ reflection high energy electron diffraction (RHEED). The intensity oscillation of the specular spot (00) testifies to the unit cell precision while the streaky pattern of the RHEED image affirms smooth surface morphology across all unit cells.

\textbf{\emph{Atomic force microscopy (AFM)}}: AFM imaging was done for the $n=2$ superlattice using a Park-systems NX10 microscope. The surface roughness and terrace heights were determined affirming the excellent film quality.

\textbf{\emph{XRD measurements}}: Post-characterization of the films were performed using both lab-based Rigaku Smartlab diffractometer and synchrotron XRD techniques. Synchrotron XRD experiments were carried out at room temperature with an incident photon energy of 15.5 keV at the Advanced Photon Source, USA, specifically at the 33-ID-D beamline. A pixel array area detector (Dectris PILATUS 100K) was employed to record and map the 2D diffraction spot for each individual $L$-scan step. Fluorescence and diffuse scattering contributions were removed through background correction, and geometric factor corrections were applied to extract the absolute intensity arising solely from the crystal truncation rod (CTR) diffraction~\cite{Bhattacharya:2025p2418490}. 

\textbf{\emph{Electrical transport measurements:}} Resistance versus temperature measurements were carried out in four probe Van der Pauw geometry using a CCR (Advanced Research Systems, USA).

\textbf{\emph{XAS, XLD and XMCD measurements:}} XAS measurements at 300 K for the Ni $L_{3,2}$, Mn $L_{3,2}$ and O $K$ edges were performed using linearly polarized light at a grazing incidence ($\sim$ 15$^{\circ}$) at the 4.0.2 beamline, Advanced Light Source (ALS), Lawrence Berkeley National Laboratory, USA. The difference was used for the X-ray linear dichroism signal to analyze the orbital symmetry ~\cite{Chakhalian:2007p1114} while the average was used for the XAS signal to determine the charge states of the cations. Absorption spectra for XMCD signal were measured under left and right circularly polarized light at a grazing incidence ($\sim$ 20$^{\circ}$) at the Ni $L_{3,2}$ and Mn $L_{3,2}$ at the BOREAS beamline, ALBA, Barcelona, Spain~\cite{Barla:2016p5144}. These spectra were recorded at 3.5 K and a magnetic field of 6 Tesla was applied parallel to the incident beam at grazing incidence. XMCD measurements were also performed at the 4.0.2 beamline of ALS, for temperature $\sim$ 79 K (Magnetic field $\sim$ 0.3 Tesla). All measurements were performed in the TEY mode.

\textbf{\emph{HAXPES measurements:}} Mn 3$s$ core level and valence band spectra were collected using a high-resolution Phoibos electron analyzer at the HAXPES beamline P22, PETRA III, DESY, Germany~\cite{Schlueter:2019p040010}. The pressure in the measurement chamber was maintained at $\sim$ 5 $\times$ 10$^{-10}$ Torr. The kinetic energy scale was calibrated using the Au 4$f$ core level spectra obtained from a gold reference sample mounted together with the films. The incident photon energy was set to 4.6 keV corresponding to an inelastic mean free path ($\lambda$) of 5.75 nm. This enables access of the entire film thicknesses as the probing depth is $\sim$ 3$\lambda$.

\section*{Acknowledgement}
The authors acknowledge the use of central facilities of the Department of Physics, IISc, funded through the FIST program of the Department of Science and
Technology (DST), Gov. of India. SM  acknowledges funding from a DST Nano Mission consortium project [DST/NM/TUE/QM-5/2019] and ANRF ARG grant [ANRF/ARG/2025/002878/PS]. NB and MB acknowledge funding from the Prime Minister’s Research Fellowship (PMRF), MoE, Government of India.   This research used resources of the Advanced Photon Source, a U.S. Department of Energy Office of Science User Facility operated by Argonne National Laboratory under Contract No. DE-AC02-06CH11357. This research used resources of the Advanced Light Source, which is a Department of Energy Office of Science User Facility under Contract No. DE-AC02-05CH11231.  We acknowledge DESY (Hamburg, Germany), a member of the Helmholtz Association HGF, for the provision of experimental facilities. Parts of this research were carried out at PETRA III using beamline P22, with beamtime allocated under proposal ID I-20231198. Funding for the HAXPES instrument by the Federal Ministry of Education and Research (BMBF) under framework program ErUM is gratefully acknowledged. Parts of this reseacrh were also carried out at the BOREAS beamline, ALBA under the proposal ID : 20250370383. We gratefully acknowledge financial support from the Department of Science \& Technology (Government of India) provided through the India@DESY collaboration.

\end{document}